\documentclass[letterpaper]{article} % DO NOT CHANGE THIS
\usepackage{aaai2026}
\usepackage{times}  % DO NOT CHANGE THIS
\usepackage{helvet}  % DO NOT CHANGE THIS
\usepackage{courier}  % DO NOT CHANGE THIS
\usepackage[hyphens]{url}  % DO NOT CHANGE THIS
\usepackage{graphicx} % DO NOT CHANGE THIS
\usepackage{natbib}  % DO NOT CHANGE THIS AND DO NOT ADD ANY OPTIONS TO IT
\usepackage{caption} % DO NOT CHANGE THIS AND DO NOT ADD ANY OPTIONS TO IT

\usepackage{booktabs}
\usepackage{multirow}
\usepackage{array}
\usepackage{booktabs}
\usepackage{longtable}

\usepackage{algorithm}
\usepackage{algorithmic}

\usepackage{newfloat}
\usepackage{listings}
\DeclareCaptionStyle{ruled}{labelfont=normalfont,labelsep=colon,strut=off} % DO NOT CHANGE THIS
\floatstyle{ruled}
\newfloat{listing}{tb}{lst}{}
\floatname{listing}{Listing}
\title{``It's OK Because...'': The Wild West of Student Rationalization \\ of AI Use in Academic Writing}
\author{
    Jiyoon Kim\textsuperscript{\rm 1},
    Kentaro Toyama\textsuperscript{\rm 2},
    Sangmi Kim\textsuperscript{\rm 2,3},
    John M. Carroll\textsuperscript{\rm 1}
}

\affiliations{
    \textsuperscript{\rm 1}Department of Information Science and Technology, The Pennsylvania State University \\
    \textsuperscript{\rm 2} Department of School of Information, University of Michigan\\
    \textsuperscript{\rm 3} Department of School of Education and Human Development, University of Miami\\
    jxk6167@psu.edu%, toyama@umich.edu,  sxk2113@miami.edu, jmc56@psu.edu
}

\usepackage{bibentry}
\begin{document}

\maketitle

\begin{abstract}
% 1 sentence background, motivation, framing
% 1 sentence overall methodology
% 3-5 sentences, key findings and conclusions
% optional 1 sentence -- conclusion, implications, future work, etc.

Generative AI challenges academic integrity not only by enabling students to delegate substantial portions of their academic work, but also by blurring the ethical boundaries by which students distinguish acceptable assistance from misconduct. Drawing on semi-structured interviews (n=20), AI chat logs, and course documents (syllabi, submitted assignments), we investigated how students themselves make moral sense of AI use in academic writing. Our analysis results in a range of novel findings: First, there are at least five distinct sites of AI-use conceptualization, ranging from faculty's intended AI policy, to students' actual AI use. Second, students use over 20 distinct rationalizations to justify AI use, such as that copying AI-generated text is victimless; that any AI text reflecting their own beliefs or their own style is their own writing; or that they are learning more by using AI -- even extensively -- than otherwise. We present a taxonomy of these rationalizations, and show how some of them are employed to justify conscious violations of course policies. Third, student rationalizations occur in both an ad hoc and post hoc manner, and they are not necessarily self-consistent. These and other findings suggest that modern AI presents a steep, ethical, slippery slope which students conceptually slide down, landing far outside the pedagogical goals and expectations of instructors. We discuss implications for educational design and AI policy.

%\textcolor{blue}{, which requries new discussions on ethical and pedagogical perspective for students' use of AI.} 
\end{abstract}

\section{Introduction}

Nearly four years after the release of generative AI (AI), significant debate over college students' use of AI in higher education remains unsettled \cite{parker2026longitudinal}. This debate is particularly consequential for academic writing, which has long served as a central mode of learning through which students develop, organize, and demonstrate higher-order thinking \cite{emig1977writing}, as it poses more challenges than reading, or listening, as it requires a complex process involving the integration of multiple types of knowledge within the limited capacity of working memory \cite{mccutchen2000knowledge}. AI itself complicates academic writing by students who can ghostwrite or easily humanize a final essay that is submitted to the course \cite{wise2024scholarly, giray2026cheating}. This causes significant challenges for instructors to assess students' essays \cite{fleckenstein2024teachers}, while students are uneasy to not use AI for various reasons like flexibility of AI, social norms, or time efficiency \cite{adnin2025examining, giray2026cheating}

%Gap Focus on cheating or how they use in different domain... coding
Extensive research has examined how college students use AI in writing \cite{jelson2026empirical}, why students are hesitant to disclose their AI use when disclosure is required \cite{fu2026everyone, adnin2025examining}, and how AI plagiarism/misconduct has emerged in higher education \cite{waqas2026understanding, bittle2025generative, balalle2025reassessing}. Moreover, recent studies have investigated how existing moral frameworks predict students' AI misconduct and cheating behavior \cite{theoharakis2025ai, hawdon2026cheating}. However, little attention has been paid to why students believe or perceive AI use to be morally acceptable including cases involving violations of course policy and how they justify such use of AI in academic writing.

To understand students' moral sense-making of AI use in academic writing, this work conducted semi-structured interview with 20 college students along with the collection of AI chat logs, writing submissions, and course syllabi. We contribute new insights into the following research questions:
\begin{itemize} 
\item How do college students interpret and negotiate course AI policies in relation to their own beliefs and actual AI writing practices?
\item How do college students make moral sense of using AI in academic writing? 
\end{itemize}

Our results suggest that expectations surrounding AI use are distributed across at least five distinct sites: faculty intention, formal policy, student interpretation, student self-policy, and student practice. Misalignments among these sites create openings for re-interpretation of (ethically questionable) AI use. We identify a very broad range of rationalizations -- at least 23 distinct rationalizations -- that college students use to justify their AI use in writing assignments in terms of harm, contribution, authorship, responsibility, and benefit. These are organized into six broader classes: victimless behavior, minimal AI contribution, ex ante contribution, post hoc contribution, responsibility denial, and perceived benefit. These rationalizations are often post-hoc, inconsistent, and flexible enough to justify increasingly extensive delegation to AI, even when students themselves recognize conflicts, tensions with course policies, or their own moral beliefs. We argue that AI policy and educational practice should move beyond articulating rules and instead help students understand the pedagogical rationale behind those policies.

\section{Related Work}
\subsection{Students Use AI in Writing}
%1. 첫째, 학생들이 AI를 실제로 어떻게 쓰는가\cite{jelson_nirvana_2026}\cite{zhang2025interplay}

%3. 셋째, 교실에서 AI 사용 기준이 어떻게 흔들리는가\cite{albers_are_2026}\cite{lin_relying_2026}

%4. 넷째, AI writing이 authorship과 oversight를 어떻게 복잡하게 만드는가\cite{draxler_ai_2024, rismani_use_2026, roe_review_2023}
AI use among students is predominantly prevalent in academic contexts, and more particularly, students use AI for academic writing beyond just grammar checking \citep{ammari2025students}. It was found very common that AI is involved in multiple phases of writing processes, including planning (asking for ideas, examples, or information), translating (turning student-provided ideas or partial text into prose) and/or reviewing (proofreading, feedback, grading, or revision of user text) \cite{jelson2026empirical}. \citet{jelson2026empirical} also found students' perform \textit{vibe-writing}, a writing mode in which students guide ChatGPT through prompts while the AI generates most or all of the essay.  

% AI use is currently prevalent \cite{jelson2026empirical, ammari2025students}. For instance, students use AI for content generation beyond just for grammar editing \citep{ammari2025students}. AI also involves in planning, translating, reviewing  \citet{jelson2026empirical} categorized students' ChatGPT queries during essay writing into four types: planning, which involves asking for ideas, examples, or information; translating, which involves turning student-provided ideas or partial text into prose; reviewing, which involves asking for proofreading, feedback, grading, or revision of existing text; and All, which involves engaging multiple writing processes at once. They further identify \textit{vibe writing} as a writing mode in which students guide ChatGPT through prompts while the AI generates most or all of the essay. The All category was the most common query type, with 54.5\% of participants using at least one All query.

Moreover, the widespread use of AI has reshaped the student perception about ownership and authorship. For example, \citet{draxler2024ai} found that people feel a greater sense of ownership when they influence AI output while using personalized AI did not have a significant impact. \citet{rismani2026use} further showed that people feel ownership when AI-generated text reflects their ideas, personal experiences, and active choices. When individuals can control AI-generated text, such as by requesting suggestions, accepting or rejecting them, or editing the AI output, they feel a stronger sense of ownership. Therefore, the ambiguity of authorship in AI-assisted writing has become a major challenge and requires urgent attention in academic policy \cite{draxler2024ai, rismani2026use}. Overall, the previous work has examined how college students write with AI in academic settings, as well as their perceptions of ownership in AI-assisted writing, less attention has been paid to why students believe it is ethically acceptable to use AI in writing.

\subsection{AI Plagiarism and Misconduct in Student Writing}
%violating standards of authorship and intellectual integrity

Historically, academic misconduct has long been a major challenge \cite{eaton2021plagiarism}. AI, however, has complicated conventional understandings of plagiarism. As AI provides information that draws from multiple sources, often lacks an identifiable origin \cite{mueller2024llms} or hallucinates by creating fake information or citations \cite{emsley2023chatgpt}, work assisted by AI may blur its ownership and/or contain inaccurate information.

These concerns have also substantially emerged in academic settings. Specifically, several types of AI-specific cheating behaviors have been witnessed in the context of writing assignments. For example, students copy and paste AI-generated text without proper disclosure, which is coined with the term \textit{AIgiarism} \cite{waqas2026understanding}. As using AI-powered writing tools, they often filter AI outputs through a tool that humanizes AI to conceal AI assistance. These form of AI ghostwriting and hidden assistance pose novel challenges for evaluating students' work properly and fairly \cite{scarfe2024real}, creating grey areas for academic integrity that fall beyond traditional plagiarism frameworks \cite{bittle2025generative, balalle2025reassessing} and raising broader questions of authorship, transparency, accountability, and learning assessment \cite{perkins2023academic, cotton2024chatting, chakravorti2025social, khosrowi2023diffusing}. 

%\citet{rismani2026use} further shows that using AI writing assistants requires active oversight, because users may accept AI suggestions without sufficiently checking whether they are accurate, appropriate, or aligned with their intended meaning. 

%\citet{roe_review_2023} similarly argue that AI-powered writing tools, including machine translators, digital writing assistants, and automated paraphrasing tools, create grey areas for academic integrity in language classrooms. 

In response, institutions have developed AI governance frameworks, revised academic integrity policies, and redesigned assessments to address AI-enabled misconduct \cite{wu2024ai, khlaif2025redesigning}. At the course level, instructors have often relied on detection-based approaches, but these tools remain limited in reliability and may increase student anxiety, produce misjudgments, and raise equity concerns \cite{ardito2025generative, liang2023gpt}. Rather than treating AI use as inherently unethical, \citet{perkins2023academic} argues that its ethical status depends on transparency and alignment with institutional expectations. Likewise, scholars have called for proactive academic integrity protocols \cite{cotton2024chatting}, while \citet{eaton2023postplagiarism} proposes that normalized human-AI writing requires academic integrity to be reconsidered in both principle and practice. These arguments indicate that what needs to be discussed is no longer merely about banning AI, but about reconfiguring academic ecosystems to support learning, authorship, and responsible human-AI collaboration \cite{slimi2026systematic}.

Yet, there is still no clear consensus on what constitutes AI cheating \cite{gruenhagen2024rapid, perkins2024decoding}. Some studies define it primarily through authorship, treating AI cheating as submitting AI-generated work as one's own \cite{perkins2023academic, chan2025students}. Others emphasize disclosure, framing misconduct as using AI assistance without permission, acknowledgment, or alignment with institutional expectations \cite{foltynek2023enai, perkins2023academic}. Assessment-focused approaches further show that the same AI use may be acceptable or unacceptable depending on task-specific rules and permitted levels of AI assistance \cite{perkins2024artificial, perkins2024decoding}. 

Furthermore, there is a gap between students and instructors regarding the extent to which they are allowed to use AI and incorporate AI-generated work into academic practices and what constitutes ethical standards. For example, \citet{lin2026relying} found agreements between students and instructors on the use of AI for ideation and revision in writing assignments, but substantial conflict existed over norms around AI-generated writing.  Another body of research suggests that AI cheating is often understood as a matter of degree. Using AI for tasks such as brainstorming, feedback, or language editing can be perceived as lower risks and harm, distinguished from other types of uses such as generating data analyses, quiz answers, and complete assignments \cite{gruenhagen2024rapid, lund2025ai, huang2025academic}. Prior research describing the emergence of novel forms of academic misconduct related to AI use highlights the importance of discussions on what constitutes acceptable and ethical academic writing practices with AI, as well as a significant gap between student and instructor perspectives. However, how and why students consider certain AI-assisted writing practices justifiable and ethical remains understudied, and understanding it is important to inform academic integrity policies and pedagogical design and assessment.  

\subsection{AI Use and Moral Framework}

% Moral disengagement theory explains how people preserve a moral self-image while violating norms by reconstructing the act, shifting responsibility, minimizing consequences, or blaming others \cite{bandura2017moral}. Similarly, neutralization theory describes how individuals reduce guilt through strategies such as denying responsibility or victimhood and condemning those who judge the behavior \cite{hawdon2026cheating}. 
Recent studies have applied existing moral frameworks to predict AI misconduct from students' moral reasoning and affect. For example, \citet{theoharakis2025ai} found that moral disengagement predicted misconduct among UK graduate business students, with stronger effects when students perceived AI as useful, used it habitually, or had stronger prompting skills. Similarly, \citet{hawdon2026cheating} found that neutralizing excuses and peer cheating were associated with AI cheating, whereas guilt was negatively associated with it. 

%Another body of work examines why students do not disclose their use of AI. AI disclosure is a situated practice shaped by environmental and social pressures, including deadlines, peer norms, emerging informal norms around AI use, and the growing normalization of AI in academic work \cite{fu2026everyone}. Yet students often avoid disclosing AI use in assignments not necessarily because they intend to cheat or deceive instructors, but because of anxiety, peer normalization, potential stigma, and ambiguity around expectations \cite{adnin2025examining}.

Recent studies have begun to examine how students justify their use of AI. For example, by using Theory of Planned Behavior (TPB) as framework, \citet{giray2026cheating} interviewed six Filipino undergraduates (who admitted to cheat in writing) and found that students often re-framed AI cheating in writing as a survival strategy rather than as academic misconduct. Their justifications varied by perceived course importance, peer normalizations, and perceived ability to evade AI detection. 

While prior studies have examined how established moral frameworks relate to students' AI misconduct, or how students admit to cheating while justifying their use of AI in writing, less is known about how students themselves define the boundary of ethically acceptable AI use in writing. %This is especially important in cases where students' AI practices violate course policies, yet students still experience those practices as reasonable, necessary, or ethically ambiguous.

%While existing studies have examined how existing moral framework is related to students' AI misconduct behavior, or study how student admit cheating but justifying their AI use in writing, less is known about how students -- themselves -- articulate ethically acceptable use of AI in writing when it includes violation of course policy, yet struggling often. %Our work addressed this gap by applying an inductive approach of understanding students' rationalizations around ethical boundaries of AI use in academic writing. 

%AI impact human Moral judgement (i.e., how AI can influence huan responsibility ) \cite{salatino2025influence}
%Writing authorship perception. People perceived still my writing eventhough New CRediT taxonomy is needed for authorship in AI-human writing \cite{draxler2024ai}%% 연속적인 기여도 측정: AI가 글을 통째로 썼는지, 아니면 사람의 아이디어를 다듬어주기만 했는지 등 기여의 정도가 다르기 때문에 이를 세분화할 체계가 필요합니다 CRediT taxonomy의 확장: 연구진은 과학 논문 기여도를 표시할 때 쓰는 CRediT 분류 체계를 수정하여, AI가 어떤 역할을 수행했는지(초안 작성, 편집 지원 등)를 구체적으로 명시할 수 있는 사용자 중심의 가이드라인을 만들어야 한다고 제안합니다.

\section{Method}

\subsection{Participants and Recruitment}
To answer our RQs, participant recruitment targeted college students who self-reported AI use for writing assignments. We processed a screening survey via Qualtrics that asked students how often they used AI for writing assignments, using a 5-point Likert scale: never, rarely, sometimes, often, and always. After processing the screening survey responses, we sent invitations to participants who selected ``sometimes'' or higher. Participants were recruited through poster flyers in campus buildings, an institutional platform for study participant recruitment (e.g., StudyFinder), and social media post (on Discord), The final sample consisted of 20 undergraduate students from 12 different U.S. universities (15 females and 5 males). 14 of them were domestic students while there were 5 international students and 1 student with dual nationality. Participants' majors were diverse, including accounting, pre-med, communication, statistics, psychology, finance, nursing, education, English and art, or cell and molecular biology. All interviews took place between December and May 2026, and participants were compensated with a \$20 Amazon gift card. 

\subsection{Data Collection}

We collected interview data with individual participants, their course syllabi, and writing assignments and AI prompts/logs. Before participating in individual interviews, participants were asked to share their AI prompts and chatlogs used to complete the assignments. We also asked students to share course policy about AI from the course syllabus or instructor's in/formal instructions and final essay submissions from the course. The submission of these information was completely voluntary, and five students shared their final essay submissions and 14 shared an AI policy for the course in the form of a syllabus. Each participant shared chatlogs via Qualtrics, so full writing sessions \footnote{one full writing session refers to one continuous conversation with an AI tool from the moment they started working on the assignment until they stopped working during that session for an essay assignment} were accessible. The submitted information was used to understand participants' AI use history and better craft questions relevant to each participant prior to the interview sessions. It was also reviewed with the participant during the interview and used to elicit more honest and vivid responses. 

Semi-structured, individual interviews were conducted virtually and lasted approximately one hour. As AI use is a sensitive topic and students are commonly hesitant to share their use \cite{adnin2025examining}, participants were informed that all information shared during the interview would be confidential and encouraged them to share their honest experiences and thoughts about AI use in writing assignments. Each interview began with questions about what the assignment was about, what their prompts aimed to do, why they asked such questions to AI, and what each participant expected AI to answer. Then, follow-up questions were asked to understand their awareness of the current AI policy within the course and their thoughts about that policy. Finally, participants were asked to share their honest thoughts on their current use of AI in terms of learning experience and ethics. The study was approved by IRB at one of the authors' institutions. %Data collection, including interviews, was conducted between December 2025 through April of 2026. 

\subsection{Data Analysis}

%After the interviews were completed, the first author coded and analyzed the transcripts, and shared with other authors with de-identified versions. 
Due to IRB restrictions on data sharing (across other authors from different institutions), only the first author and last author had access to the students' raw chat logs. Following discussion with and advice from the IRB liason, the first author de-identified all participants' personal information, such as names, email addresses, and course names, before sharing the students' chat logs, or interview transcripts with other research team members, replacing identifiers with a Participant ID (PX). This constraint shaped our analytic workflow. First author conducted hands-on coding of raw data, while the all team members participated in interpretation, refinement, and consensus-building over de-identified data.

We conducted thematic analysis, allowing codes and themes to emerge from the data without a predetermined coding framework \cite{clarke2017thematic}. First, the first author reviewed interview recordings and transcripts multiple times and identified portions of the interview transcripts in which participants described or justified their use of AI in academic writing. Using these excerpts as units of analysis, we went through several rounds of iterative thematic clustering, beginning with an initial organization of participant utterances around recurring forms of ethical reasoning. As we compared excerpts across participants, it became clear that participants were not simply describing whether they used AI, but repeatedly offering different moral justifications for why particular AI-assisted writing practices were acceptable, necessary, harmless, or ambiguous. We therefore focused our analysis on the rationalizations participants used to make sense of their AI use, rather than on classifying AI-use practices alone.

Through this process, we initially clustered individual excerpts into 46 subcodes that captured recurring patterns in participants' rationalizations. These subcodes were then consolidated into 23 rationalization categories based on shared similarities, which were further grouped into six broader rationalization classes: victimless behavior, minimal AI contribution, ex ante contribution, post hoc contribution, responsibility denial, and perceived benefit (see Table~\ref{tab:rationalizations-part1}). Definitions of each rationalization were developed inductively from participant accounts and refined through discussion among all research team. Due to space constraints, we report the 23 category level with six classes in this paper.

At that point, data were analyzed in two complementary ways. First, iterative thematic clustering continued in order to identify higher-level patterns across participants' moral reasoning. Second, excerpts that reflected ethical reasoning about AI use were coded according to rationalization categories wherever sufficient information was available. Throughout all cycle, research team met regular basis to compare interpretations, discuss similarities and differences across rationalizations, refine category boundaries, and resolve disagreements through consensus. This process resulted in 23 student rationalizations as well as the broader findings regarding students' flexible, post-hoc, and sometimes internally inconsistent moral sense-making around AI use in academic writing.

\section{Findings}

%some students ask if this is confidenctial 
%AUstin even write letter to faculty
%p15 sanvvi ethically wrong - still do it. 
%you have to know literuate ,, you know it well......
% literature -, whats new discussion -  from 

\subsection{Five Sites of AI ``Policy''}

Our analysis of syllabi and interviews together suggests that for any given course, there are at least five distinct ``sites'' that house concepts of appropriate AI use for academic writing: (1) \textit{\textbf{Faculty Intention}}, instructor expectations of student AI use; (2) \textit{\textbf{Formal Policy}}, the explicitly specified rules, usually written down in syllabi or assignment instructions; (3) \textit{\textbf{Student Interpretation}}, a student's own understanding, recall, and/or interpretation of formal policy, (4) \textit{\textbf{Student Self-Policy}}, a student's own normative views about AI use; and (5) \textit{\textbf{Student Practice}}, a student's actual use of AI. 

We find that the AI ``policies'' -- whether explicit or implicit, formal or informal, specific or vague -- that are expressed in each of these sites often differ from the others (though, not necessarily). For example, while we did not interview faculty directly, it was clear from syllabi and participant interviews, that Faculty Intention (Site 1) was not always clearly expressed in their course's Formal Policy (Site 2). One participant (P8) explained that they could copy and paste AI-generated text because the course syllabus allowed AI use for idea generation. Whether the instructor intended to permit direct copying from AI output was unspecified in the syllabus, though presumably, they had some intention about it. Relatedly, A faculty co-author of this paper notes that in their own teaching, it took several iterations of a course to converge on a well-specified written policy that expressed their expectations of AI use -- in other words, their first attempts did not perfectly match their intention. 

%A formal policy in syllbi did not always translate into students' clear understanding of that policy, and students' own beliefs about reasonable AI use did not always align with either the written policy or their actual practices. 

Similarly, some students misremembered their course AI policies, or mixed up policies between courses, demonstrating the distinction between Formal Policy (Site 2) and Student Interpretation (Site 3). Many of our participants expressed that they knew what the course policy was (Student Interpretation, Site 3), but applied a policy that made more sense to them (Student Self-Policy, Site 4). For example, some participants suggested that strict AI prohibitions were unrealistic or educationally unhelpful, even when they recognized that instructors might disapprove of their use. And finally, we saw several instances where students violated their own stated norms (Student Self-Policy, Site 4) in actual practice (Student Practice, Site 5). 

It might seem obvious that AI use is shaped not only by faculty expectations and formal rules, but also by students' interpretations, personal beliefs, and ultimate practice. Yet, the fact of these five sites suggests that AI presents myriad opportunities for misinterpretation, second-guessing, confusion, fudging, and misconduct.

\subsection{The Diversity of Moral Justifications for AI Use}
%\kt{Begin with: ``We identified 24 (???) distinct types of moral justifications mentioned by our participants regarding their AI use.}\jk{Got it -- Thank you so much!!!!}
%\jk{All Classes/RX should be re-checked before final submission +Participants' AI practice  }

We identified 23 distinct moral justifications that participants mentioned regarding their AI use -- henceforth ``Rationalizations'' (R1--R23) These 23 Rationalizations can be grouped into six higher-level categories, which we call \textit{Rationalization Classes} or simply, \textit{Classes} (C1--C6). Table~\ref{tab:rationalizations-part1} presents our taxonomy, along with representative participant quotes for each Rationalization. Space does not permit an in-depth description of every Rationalization, but below, we describe each Rationalization Class and offer examples of the Rationalizations that it comprises \footnote{Participants' AI writing practices varied, ranging from making outline, paraphrasing AI-generated text, and expanding self-written texts to copy-paste AI-generated text into submitted assignments. Our taxonomy classifies the rationalizations participants used to justify these practices, rather than the practices themselves. Thus, the same practice could be justified through different rationalizations, and the same rationalization could support different levels of AI involvement.}.

The first Class, \textit{\textbf{C1: Victimless Behavior}}, involves two Rationalizations in which participants suggest their AI use is acceptable, because no (human) person is harmed by it, or because the only person being harmed was the participant themselves. Some participants implied that plagiarism was bad only because it hurt others, but as P16 noted, \textit{``AI doesn't have a soul... giving AI credit [or] respect as if AI had ownership or legality to certain information... that's simply not true.''} A few participants stated directly that copying and pasting AI-generated text did not harm other human beings and therefore did not raise an ethical problem.
%claims related to \textit{R1: No Human Victim} and \textit{R2: AI-Synthesized Sources}. Under \textit{R1: No Human Victim}, participants justified AI use by arguing that plagiarism and authorship norms are primarily designed to protect human creators, human labor, or human originality. Because AI was not viewed as a person (e.g., \textit{``AI doesn't have a soul...giving AI credit \ldots respect as if AI had ownership or legality to certain information it shared when that's simply not true (P16)''}), and because AI-generated text was not seen as belonging to a human author in the same way as published writing, participants reasoned that using AI output did not constitute a serious ethical violation. For example, one participant contrasted AI use with plagiarism, which they described as \textit{``either copying or like stealing credit from the actual people (P12)''}.
Similarly, some participants argued that AI synthesizes information rather than copying from a specific source, thus eliminating any human victims of plagiarism. Interestingly, one participant noted that there was no harm to instructors, and went as far as to suggest the only possible harm as being to oneself: \textit{``It's not hurting anybody \ldots I think you are hurting yourself more than you're hurting like a teacher or other people (P6)''}.

The second Class, \textit{\textbf{C2: Minimal AI Contribution}}, involves Rationalizations where AI's contribution is cast as trivial or equivalent to other allowable forms of assistance, thereby limiting its moral significance. 
%related to \textit{R3: Busywork}, \textit{R4: Facts Only}, and \textit{R5: Like Other Allowable Support}. Under this class, participants minimized AI's role by portraying its contribution as trivial or equivalent to other allowable forms of assistance, thereby limiting its moral significance.
This Class includes \textit{R3: Busywork}, the most common Rationalization raised by participants. They argued that AI was OK for ``busywork,'' because such assignments were devoid of learning value, unimportant, meaningless, or low-stakes (unlike, e.g., exams). A related Rationalization was \textit{R5: Like Other Allowable Support}, where participants claimed that AI was similar to traditional tools or forms of support, such as human editors, grammar checkers, or writing centers, and therefore also acceptable to use. For example, one participant compared AI to receiving suggestions from a human writing coach, even though the text that expanded the idea was generated by ChatGPT: \textit{``If I don't have that answer for myself, then she [the coach] would say, you could do this. That's like, the same as what ChatGPT is doing for me.''} (P13).
%They also described these assignments as not relevant or helpful to their future careers, or would help managing work--life balance (e.g., \textit{``[I would rather] spend time with my girlfriend... (P13)''} ). Some participants also argued that using AI allowed them to allocate more time to more important work, such as final exams or exams related to their major.
%Under \textit{R4: Facts Only}, participants characterized AI-generated content as common knowledge or fact-based material rather than personal opinion or creative writing. 
%Under \textit{R5: Like Other Allowable Support}, they claimed that AI was similar to traditional tools or forms of support, such as editors, writing centers, or Google, because these resources help students develop or expand on their own ideas. For example, one participant compared AI to receiving suggestions from another source: \textit{``If I don't have that answer for myself, then she would say, you could do this. It's like, the same as the ChatGPT is doing it for me (P14)''}.

The common logic across this Rationalization Class was that AI's contribution to the final work was not substantial enough to threaten authorship or academic integrity. Rather than denying that AI had contributed, participants reframed its use as minor assistance. 

%\jk{Based on Kentaro's comment in google doc, this class (R3) and (R4) seems to be changed a lot. I will get back to this.}

The third Class, \textit{\textbf{C3: Ex Ante Contribution}}, involves Rationalizations %related to \textit{R6: My Ideas}, \textit{R7: My Directions}, and \textit{R8: My Curation}. This class captures cases 
in which participants provided some core content before turning to AI and emphasized that this prior input was the critical part of the work.
These rationalizations suggest that AI use is acceptable because the core idea, task direction, or source materials came from the participant. For example, Rationalization \textit{R6: My Ideas} involves cases in which participants described already having ideas in mind but needing help to articulate or to expand on them. One participant explained that they had \textit{``ideas in my mind, but it's hard to articulate''} (P3), while another said, \textit{``I have the story in my mind, but I really wanted [AI] to expand it for me''} (P14). Similarly, 
Rationalization \textit{R7: My Directions} refers to cases in which participants explained that their provision of detailed prompts to AI was the essential work. As one participant explained, \textit{``I'm the person who has come up with the prompts and the AI has come up with the essay'' (P13)}. %Participants also emphasized their role in curating important resources that shaped the AI's response, which we refer to as \textit{R8: My Curation}.

With Class C3, the student sees themselves as the primary originator or director of the work, and therefore creatively and ethically above board in using AI. Like the previous Class C2, Minimal AI Contribution, this Class also discounts AI's role, but here, the Rationalizations prioritize the student's substantive and ideational contributions as primary and essential. %the which is only providing  to . In these cases, AI is framed as a tool that helps articulate, elaborate, or direct the output, while participants treat their own prior input as the morally relevant contribution.

%KT: One possibility is to change R3 and R4 in the following way: R3 is when the user provides some core content before turning to AI; R4 is when the AI provides the core content, but the user then agrees to it.
%It might be that I7 is "My Ideas, Which I Originated" and this one is "My Ideas, Which I Recognize When AI Generates Them."

In contrast to C3, the fourth Class, \textit{\textbf{C4: Post Hoc Contribution}}, involves Rationalizations in which the student's contribution occurs after AI has output substantive content, and often only stylistic or confirmatory ways. 
%related to \textit{R9: My Style???}, \textit{R10: Selective Use}, \textit{R11: My Paraphrasing}, and \textit{R12: My Verification}. In this class, participants justified AI use when AI provided the core content, but they subsequently accepted, agreed, or aligned self with its output. This class is distinct from \textit{User-Initiated Contribution}: the latter involves ideas that participants brought to AI beforehand, whereas this class involves AI-generated content that participants later treated as personally attributable because it expressed something they agreed with, could imagine saying themselves, or felt was meaningfully aligned with their own style and perspective. 
Thus, with Rationalization \textit{R10: My Paraphrasing}, students argue that paraphrasing AI output makes it their own; or, with Rationalization \textit{R12: My Verification}, that checking AI output for accuracy was sufficient. 

Most interesting in this Class is Rationalization \textit{R12: My Style}, in which participants explain that if the output ``sounds like'' them, it is as good as if they wrote it. Some participants described either prompting or experiencing AI take on their writing style. One participant explained that she could even ask AI to generate prompts that imitated her thinking style:
\begin{quote}
\textit{``It has recognized a pattern of my way of writing and how I express things in, like projects or emails or anything like that. So now, it knows what kind of person I am like in the way of writing. So now, it gives the prompts similar to like, what my thinking process is''} (P9).
\end{quote}
%Participants also attributed AI-generated text to themselves when they paraphrased it in their own words, reasoning that the text became their own through this transformation. Finally, participants justified AI use as ethically sound when they actively verified the accuracy of AI output, checked for errors, traced claims back to original sources, and cited those sources.
%Altogether, this class suggests that participants claimed ethical ownership over AI-generated material through post-output practices. %By endorsing, personalizing, paraphrasing, or verifying AI-generated content, participants reasserted control over material that had initially been produced by AI.

The fifth Class, \textit{\textbf{C5: Responsibility Denial}}, %involves Rationalizations related to \textit{R13: Agency Denial}, \textit{R14: Normalization}, \textit{R15: Inevitability}, \textit{R16: Instructor Indifference}, \textit{R17: Normlessness}, and \textit{R18: No Consequences}. This class captures
comprises Rationalizations that weaken, displace, or externalize participants' sense of responsibility.
With Rationalization \textit{R13: Normalization}, participants pointed to external norms, arguing that AI use had become socially normalized and practically unavoidable. Several %described competitive pressure, expressing concern that not using AI would mean \textit{``putting yourself in a very disadvantaged position''} (P13). Others 
explained that they needed to use AI to keep up: %described a sense of unfairness when they perceived that their peers were using AI: 
\textit{``[If I am] the only one writing my own work and everyone else is using ChatGPT, I feel like they're ahead of me because they just saved four hours.''} (P18) Some participants also raised fairness concerns, especially when classmates appeared to use AI and received good grades: \textit{``Some of my classmates who just throw it [into AI], generate it, don't even check it, [and] they get higher grades than I do''} (P17). Participants also invoked \textit{R15: Instructor Indifference} and \textit{R17: No Consequences} as justifications. For example, some participants blamed instructors, arguing that assignment instructions were ambiguous or difficult to understand: \textit{``That is faculty's fault. What can we do when they are not helping?''} (P3). 
%one participant argued that faculty use of AI made student use seem more acceptable: \textit{``I see a lot of my professors use AI to generate [projects]. So I'm like, if they're using it, why can't I? (P19)''}. 
Others emphasized the absence of consequences, suggesting that AI use felt acceptable because it was unlikely to be detected or punished. As with other Rationalizations in Class C5, participants shifted responsibility away from themselves or denied any wrongful intention. %by either framing AI as the primary actor, particularly in relation to plagiarism, or by describing their own actions as unintentional. For example, when one participant claimed that, if anything was plagiarized, \textit{``AI is the one plagiarizing,''} they were not merely describing AI's technical role; rather, they were relocating accountability from the student to the system. Similarly, another participant claimed that AI use would be ethically acceptable \textit{``if [a] few sentences [were] copy pasted from AI unconsciously (P11)''}.
%

%This rationale differs from User Contribution and Control because it does not necessarily require substantial editing, verification, or curation. Instead, the participant treats alignment between the AI output and the self as enough to make the output feel personally attributable.

The sixth and last Class, \textit{\textbf{C6: Perceived Benefit}}, %involves claims related to \textit{R19: Time Economy}, \textit{R20: Educational Value}, \textit{R21: Better Writing}, and \textit{R22: Learning-oriented Intention}. This class 
justifies AI use through its practical, educational, or performance-related advantages. Many participants, for instance, invoked Rationalization \textit{R19: Time Economy}, stressing that AI use saved them time. One participant went as far as to say, \textit{``[I would rather] spend time with my girlfriend...''} (P13). Notably, this Rationalization was frequently paired with Rationalization R3: Busywork. 

Some participants viewed paraphrasing AI-generated text as a way to preserve learning, reasoning that they still learned by rewriting the content in their own words (\textit{R20: Educational Value}). Others perceived AI-generated writing as higher quality than their own writing and argued that this could support better learning (\textit{R21: Better Writing}). One unique participant (P6) claimed that AI use was ethically acceptable if the user had some degree of sincere intention to learn (\textit{R22: Learning-oriented Intention}). Participants also justified AI use by emphasizing positive outcomes or performance benefits, such as receiving a good grade, earning faculty praise, or simply completing the assignment (\textit{R23: Better Outcome)}. 
%This class foregrounds the perceived benefits of AI use. Participants treated these practical, learning value, and performance-related advantages as reasons why AI use could be acceptable.
In all of the Rationalizations in Class C6, participants reason that AI use's beneficial effects make it permissible -- the ends justified the means. 

%Notably, participants often connected this Rationalization Class to time management: Using AI allowed them to re-allocate time to more important work, such as final exams or courses in their own major.  

%R7 generation based: This rationalization is generationally bounded;  it does not apply to younger cohorts whose learning occurred during AI's presence.

% Required package:
% \usepackage{booktabs}

\begin{table*}[]
\centering
\scriptsize
\setlength{\tabcolsep}{3pt}
\renewcommand{\arraystretch}{0.90}
\caption{Taxonomy of Student Rationalizations for AI Use in Academic Writing (Part 1 of 2).}
\label{tab:rationalizations-part1}
\begin{tabular}{@{}p{0.28cm} p{2.25cm} p{3.25cm} p{9.45cm}@{}}
\toprule
\textbf{\#} & \textbf{Rationalization} & \textbf{Definition} & \textbf{Representative Quote} \\
\midrule
\multicolumn{4}{@{}l}{\textit{\textbf{Rationalization Class C1: Victimless Behavior}} --- ``AI use has no human victim; ethics are inapplicable.''} \\
\midrule

R1 & No Human Victim & 
It is OK because plagiarism and authorship norms exist to protect human authors, human effort, and human ownership. Since AI is not a person, does not exert human-like effort, and no human is directly harmed, AI-generated text lacks a morally relevant victim. & 
%``AI doesn't have a soul. So AI doesn't have a spirit and a soul. So a person has original thought and a person has creativity and original thought... Ultimately, AI is not a person, and giving AI credit for like, plagiarizing saying, well, I need to credit AI. That would be giving credence to AI and respect as if AI had ownership or legality to certain information it shared when that's simply not true'' (P16)
%\newline
``It wasn't written by another human... it wasn't somebody's idea that I stole... So I guess it's fine'' (P11) \\
\cmidrule(l){3-4}

R2 & AI-Synthesized Sources & 
It is OK because AI synthesizes and rewrites information rather than copying from a specific source, and since I can cite AI-generated text, I do not see it as plagiarism. & 
``With ChatGPT, [it] is not copy pasting from an external source. What it's doing is... analyzing the information from four or five different sources, then clubbing it into one, writing it as a whole different thing... So this is not plagiarism, because this is not directly copy paste [from] the sources'' (P9) \\

\midrule
\multicolumn{4}{@{}l}{\textit{\textbf{Rationalization Class C2: Minimal AI Contribution}} --- ``AI contribution is too minimal to be seen as an ethical issue.''} \\
\midrule

R3 & Busywork & 
It is OK because AI contribution to only ``busywork.'' & 
``With a lot of these assignments, they're not particularly... challenging, but it's just a lot of busy work... it's just not worth my time to actually... consume all of that content...'' (P2) \\
\cmidrule(l){3-4}

R4 & Facts Only & 
It is OK because factual information cannot be owned. & 
``When you're writing, you can use like the facts of it'' (P17) \\
\cmidrule(l){3-4}

R5 & Like Other Allowable Support & 
It is OK because AI's role is the same as existing writing resources and tools such as the writing center, proofreading editor, Grammarly, and Google. & 
``The writing center does the same thing... she's also... asking me to think, but it also helping me, like a booster... It's like, the same as the ChatGPT is doing it for me'' (P14) \\
\cmidrule(l){3-4}

%R6 & Like Other Allowable Output & 
%It is OK when AI produces the same wording, information, conclusion, or result that would have emerged anyway. & 
%``You're gonna get the same information, whether it's Google or ChatGPT. Then maybe fine'' (P8) \\

\midrule
\multicolumn{4}{@{}l}{\textit{\textbf{Rationalization Class C3: Ex Ante Contribution}} --- ``I take care of the critical parts and exert control over the AI processes and outputs.''} \\
\midrule

R6 & My Ideas & 
It is OK because the core ideas, intentions, or thoughts originate from me. AI helps articulate, expand, or clarify what I already had in mind. & 
``When I get a topic in my head, at that point, I know what the topic was. I know what it needs... but at that point it's like jumbled up in my head. I can't really put out... an organized statement... So [AI] organized my thoughts in a better way'' (P10) \\
\cmidrule(l){3-4}

R7 & My Directions & 
It is OK because I direct AI through prompts. & 
``I'm the one [who] gave it a prompt. So I already dictated for it what it should give me, what I need... and how I want the information to be dispensed back to me'' (P7) \\
\cmidrule(l){3-4}

R8 & My Curation & 
It is OK because I do the research or select relevant materials before using AI. I select what information matters before AI performs the writing or organizing task. & 
``I give it the assignment instructions and the necessary resources. So for example, if it's a video documentary, I give it an AI summary of that documentary. If it is a book or an article... I just attach the PDF. If it's a picture, I attach the image... I give it the exact instructions and all of the resources it needs'' (P2) \\
\cmidrule(l){3-4}

\midrule
\multicolumn{4}{@{}l}{\textit{\textbf{Rationalization Class C4: Post Hoc Contribution}} --- ``I revise or check AI's work afterward.''} \\
\midrule

R9 & Selective Use & 
It is OK because I do not take AI text wholesale. I take only portions of it. & 
``I guess that would be fine. Like, one or two sentences [copy-pasted] is fine. I guess it's not like latant copying'' (P11) \\
\cmidrule(l){3-4}

R10 & My Paraphrasing  & 
It is OK because I transform AI output through editing, addition, or iterative refinements. & 
``I basically just paraphrase it and kind of condense it... I just go sentence by sentence, like condensing and paraphrasing each sentence in my own words for the assignment'' (P2) \\

\bottomrule
\end{tabular}
\end{table*}

\begin{table*}[p]
\centering
\scriptsize
\setlength{\tabcolsep}{3pt}
\renewcommand{\arraystretch}{0.90}
\setcounter{table}{0}
\caption{Taxonomy of Student Rationalizations for AI Use in Academic Writing (Part 2 of 2).}
\label{tab:rationalizations-part2}
\begin{tabular}{@{}p{0.28cm} p{2.25cm} p{3.25cm} p{9.45cm}@{}}
\toprule
\textbf{\#} & \textbf{Rationalization} & \textbf{Definition} & \textbf{Representative Quote} \\
\midrule
\multicolumn{4}{@{}l}{\textit{\textbf{Rationalization 4: Post Hoc Contribution (continued)}}} \\
\midrule

R11 & My Verification & 
It is OK because I verify, fact-check, and cite the original sources. & 
``I'm actually going back to like, check it and make sure everything is like, correct'' (P17) \\
\cmidrule(l){3-4}

R12 & My Style & 
It is OK to use AI if it sounds like me. & 
%``It's [AI-generated text is] still words I agree with and I believe in'' (P6)
%\newline
``I have spent a lot of time training my ChatGPT to sound like me and like to give good responses, so whenever [I] put in [a prompt], I'm confident that it's going to be sounding like me'' (P9) \\

\midrule
\multicolumn{4}{@{}l}{\textit{\textbf{Rationalization Class C5: Responsibility Denial}} --- ``There is nothing wrong with me using AI because it has now become common and no one cares.''} \\
\midrule

R13 & Normalization  & 
It is OK because everyone uses AI. It is normal. & 
``It's not like I'm the only one doing it... but every other person in my class, same thing.'' (P18) \\
\cmidrule(l){3-4}

R14 & Inevitability & 
It is OK because AI use is inevitable. & 
``We already have to accept this is a technology that is part of us, and as time goes by, it is continuing to develop, and generally, those who don't use it will be left behind'' (P13) \\
\cmidrule(l){3-4}

R15 & Instructor Indifference & 
It is OK because faculty do not care, give vague instructions, or implicitly permit AI use. & 
``They [faculty] don't really care about [assignments]... they don't even look at it... they never write notes... never give us feedback... they only worry about our exams'' (P18) \\
\cmidrule(l){3-4}

R16 & Normlessness & 
It is OK because there are no clear rules about AI use yet. & 
``AI is so Wild Wild West... there's not a lot of rules surrounding it... since it's such a new technology, we haven't really grasped the concept'' (P19) \\
\cmidrule(l){3-4}

R17 & No Consequences & 
It is OK because there are no consequences, such as unlikely to be detected, punished, or grade penalized. & 
``It feels like a guilty conscience, but I'm still submitting it because I know that I'm not going to get in trouble'' (P15) \\
\cmidrule(l){3-4}

R18 & Agency Denial & 
It is OK because the action is not really mine; either AI performed it, or I did not consciously or intentionally act. & 
``If [there is] anything AI plagiarizes, AI is the one plagiarizing, because it's getting information from the Internet'' (P16)
\newline
%``If a few sentences are copy-pasted from AI unconsciously, that should be fine'' (P11) 
\\

\midrule
\multicolumn{4}{@{}l}{\textit{\textbf{Rationalization Class C6: Perceived Benefit}} --- ``Using AI is beneficial for me.''} \\
\midrule

R19 & Time Economy & 
It is OK because of time and effort considerations ranging from convenience to necessity, such as overwhelming workload or language barriers. & 
``Instead of spending four hours filling out this paperwork, it kind of takes 30 minutes'' (P18) \\
\cmidrule(l){3-4}

R20 & Educational Value & 
It is OK because I still learn the content through AI use. & 
``When I put into AI summary, like while I'm [paraphrasing AI's essay sentence by sentence], I still learn the content. Like while I'm writing it, since I am typing it out, the information from the documentary and the information from the book that I should have read about is still going into my brain because I'm reading this stuff from the AI [essay], and I'm [re]writing the essay'' (P2) \\
\cmidrule(l){3-4}

R21 & Better Writing & 
It is OK because AI is a better writer than me. & 
``I feel like I'm more confident when ChatGPT writes it for me because it has better approaches to grammar and transitioning phrases and structuring'' (P9) \\
\cmidrule(l){3-4}

R22 & Learning-oriented Intention & 
It is OK when my intention is to learn. & 
``It's a lot about the intentions that you have when you're using it. If you're trying to just get the class done and you don't really care, then obviously you're hurting yourself and learning less, and that's more unethical... But if you are trying to get something out of it... then it's more ethical'' (P6) \\
\cmidrule(l){3-4}

R23 & Better Outcome & 
It is OK because the outcome is good, such as a better grade or faculty approval. & 
``Before AI, my essays were terrible... now it's easier to actually go for an A than just go for a C'' (P12) \\

% \midrule
% \multicolumn{4}{@{}l}{\textit{\textbf{Rationalization Class C7: Agency Denial}} --- ``I am not the actor.''} \\
% \midrule

\bottomrule
\end{tabular}
\end{table*}

\subsection{Belief vs Behavior Mismatch}

%\jk{1. Wrong and distressed, but continued}
   %= 잘못이라는 걸 알고, 실제로 괴로워하지만 계속 사용함

%\jk{2. Wrong, but pragmatically set aside}
   %= 잘못이라고 생각하지만 시간, 시험, 귀찮음 때문에 윤리적 우려를 잠시 접어둠

%\jk{3. Not wrong, but still hides/manages traces}
   %= 본인은 문제없다고 말하지만, 실제로는 disclosure를 줄이거나 AI detector를 돌리는 등 숨기려는 행동을 함

We have noted already that participants' ideas about appropriate AI use, whether based on Student Interpretation of policy or Student Self-Policy, can diverge from actual Student Practice. %Here, we note that these instances occurred frequently in our data, and that the moral reactions among participants can be described as one of three: (1) Some participants viewed their AI use as wrong and expressed moral distress; (2) some recognized ethical concerns but set them aside without overtly expressed struggle; and (3) some did not see a moral problem, though they acknowledged a technical discrepancy. 

%\paragraph{Wrong and distressed, but continued use.}
%The first form of mismatch occurs when 
Some participants recognized that their AI use violated course rules or their own moral belief, yet continued to use AI while experiencing genuine moral distress. P16, for example, expressed a strong desire to follow faculty rules and act ethically, stating, \textit{``I want to do what my teacher is asking me to do. I want to obey her, and I want to do what's right''} (P16). Yet P16 also described continuing to use AI despite a strict course policy that prohibited it (e.g., \textit{``[faculty is like] never to use AI. Ever, never, ever, ever,  ever, ever. She's like, don't look at it, don't smell  it, don't touch it''} (P16)) .  Rather than helping them avoid AI, the prohibition intensified their internal moral conflict:
\begin{quote}
\textit{``I don't think that is actually helping me stay away from it, and it's actually creating conflict for me, because I'm breaking the rules.''} (P16) 
\end{quote}

P16 summarized this distress by saying that the policy was not helping them avoid AI; rather, \textit{``all it's doing is creating moral conflict''} (P16). This case shows that continued AI use was not always driven by indifference or lack of ethical awareness. In some cases, participants were highly aware of the ethical stakes and experienced significant distress precisely because their practices conflicted with their own moral commitments.

%\paragraph{Wrong, but unconcerned.}
Some other participants acknowledged that a practice was ethically problematic but treated that concern as secondary to immediate pressures such as time, workload, exams, or convenience. For example, P15 described copy-pasting AI-generated text as \textit{``not very ethical,''} but explained that they had to focus on a major exam and therefore \textit{``put aside [their] ethical concerns''} (P15). Similarly, another participant framed the decision as a tradeoff between morality and time:

\begin{quote}
\textit{``In this paper, I didn't follow my moral compass and at the end, I just copy pasted the conclusion... It depends how much I'm willing to balance my morals. There's like balancing my time as well.''} (P19)
\end{quote}

These participants did not necessarily describe prolonged moral distress. Instead, they framed ethical concern as one consideration among others, which could be overridden by competing academic or practical demands.

%\paragraph{Not wrong, but manage its visibility.}

For some participants, AI use was not morally problematic, but nevertheless behaved as if it needed to be hidden or made less visible. For example, some participants argued that paraphrasing AI-generated text made the final work their own, yet still ran drafts through AI detectors before submission to reduce the chance of being caught. In these cases, participants' stated belief that their work was ethically acceptable coexisted with practical efforts to manage detection risk.

One participant explained that, even when an instructor's policy allowed AI use if students disclosed their prompts, they did not feel comfortable disclosing the full extent of their AI use. Instead, they disclosed only the first few prompts because they believed that attaching the full prompt history would lead the instructor to judge them negatively:

\begin{quote}
\textit{``I don't feel like I can attach like 25 screenshots of my ChatGPT prompt that [faculty] would feel crazy... Even if a faculty says that, oh, you know, just attach a prompt, I don't think they will be happy with us ''} (P5).
\end{quote}

A few students claimed that, as long as they did not copy and paste AI-generated text, using AI was ethically acceptable. However, when their final essays were compared with their AI chat logs, it appeared that they had copied and pasted AI-generated text; two students were not aware that this was what they had done in practice.

%Together, these three patterns show that participants' moral reasoning about AI use was not simply a matter of whether they believed AI use was right or wrong. Some participants experienced genuine moral distress, others pragmatically set aside ethical concerns, and still others claimed that AI use was acceptable while managing how visible that use became. In all three cases, students' stated beliefs, course policies, perceived instructor judgment, and actual AI practices did not fully align.
 
\subsection{Back-and-Forth Contradictions Within the Person}

Thus far, we have analyzed participant rationalizations of their AI use in isolation. But, our interview transcripts also exhibit many instances of torturous routes of moral reasoning where rationalizations often evolved during the interview; rarely did students start with a single rationale and remain with it. Their thinking was fluid, unfinished, and sometimes internally contradictory. Among the 20 participants, 15 exhibited this pattern. 

We present one participant as vivid example of back-and-forth rationalization due to space limits: %P18, who rationalized AI use externally through normalization, peer behavior, policy irrelevance, and fairness, and 
P6's reasoning moved among \textit{R3 Busywork}, \textit{R6 My Ideas}, \textit{R10 My Paraphrasing}, \textit{R19 Time Economy}, and \textit{R22 Learning-oriented Intention}. Yet P6 repeatedly experiencing moral discomfort and conflict with course policies.

\textbf{P6 example.} P6 shows ethical discomfort repeatedly but then softened through distinctions about multiple rationalizations. When asked about their own AI use, P6 admitted that using AI, including copy-pasting AI-generated text, could be unethical when course policies prohibit such use or require disclosure. However, P6 also explained that they often do not disclose their use because it is \textit{``really hard for [faculty] to prove.''} Still, P6 acknowledged that this way of using AI is ethically problematic:

\begin{quote}
\textit{``Honestly, it’s probably not very ethical. I do a lot of my homework with AI, and it doesn't always mean I’m learning.''} (P6)    
\end{quote}

P6 then distinguished between more and less ethical uses of AI depending on the type of task and the degree of personal involvement. Writing assignments, for example, were perceived as more ethical because P6 still had to align the AI-generated text with their own thoughts and revise it into something they agreed with (\textit{R6 My Ideas}):

\begin{quote}
\textit{``AI helps me a lot, because I can't think about what I want to say. But if I have AI, write stuff and then I change it to say more what I want to say, I still think that's ethical, because it's still words I agree with and I believe in''} (P6)  
\end{quote}

And, P6 also explained, (\textit{R10 My Paraphrasing}): \textit{``more of it will end up being my original work by the end, because I will change it.''} 

At the same time, P6 acknowledged that failing to follow course AI policies, such as citing where AI was used, would be unethical. They described this as \textit{``unethical, which would be if I didn't cite the AI, if I just use the work as my own, just like plagiarism.''} P6 then stated the contradiction even more explicitly: \textit{``if they just say specifically not to use AI and I do it and I don't cite it, I think that's definitely unethical, even though I still do it.''} They added, \textit{``if I have a teacher that doesn't want me to use AI at all, I'm still gonna do it, but I'm lying to them now, and that's wrong.''}

Yet P6 then shifted again, this time minimizing the importance of certain classes (\textit{R3 Busywork}): \textit{``I use AI a lot more in a class that I just want to pass and I don't really care about it.''} P6 also reframed AI use through intention, suggesting that using AI could be more ethical when the goal was not simply avoidance, but saving time or getting some value from the task (\textit{R22 Learning-oriented Intention} and \textit{R19 Time Economy}). As P6 put it, \textit{``for me, it's a lot about the intentions that you have when you're using it.''} They further explained that \textit{``if you are trying to get something out of it, you're trying to save yourself time, though, then it's more ethical...''}

P6 continued to wrestle with whether particular uses of AI were more or less ethical by focusing on whether they had added their own thoughts. For example, when discussing online discussion posts, P6 described submitting AI-generated text without reading the textbook as \textit{``probably less ethical because I didn't read the textbook... I didn't really include any of my own thoughts.''} In this way, P6 did not treat AI use as simply ethical or unethical. Instead, they evaluated it along a spectrum based on disclosure, task value, intention, and how much of their own thinking was incorporated.

P6 also expressed concern about the longer-term harms of relying on AI. Because their dependence on AI had become significant, P6 worried that it was eroding their ability to articulate thoughts in real-life conversations: \textit{``I think that I’m losing some of the skills to talk to other people... it’s hurt me.''} Yet immediately afterward, P6 explained why the practice nonetheless continues:

\begin{quote}
\textit{``it is still hurting me... But again, when I use it in the short term, I don't think too much about that long term effect. I just think about, I just want this assignment done, and that’s it.''} (P6)
\end{quote}

%slippery slope : AI use evolved over time "I used to ask it to write me an outline, and then I would write the actual essays... Now I have a busier schedule, and I'm also just more lazy, I think, but I just make it write most of the words" (P6)
%P13 , P12

%\subsection{Post-Hoc Rationalization}
%\kt{I think the subsections above and below this line can be merged; so, I did. ??? Cut this comment after reading.}

One can infer from the above that college students' moral rationalizations of AI use are often post-hoc. Participants had not thought through the morality of their situations in advance. They often contradicted their multiple rationalizations and misjudged the relationship between their stated beliefs and actual behavior about appropriate AI use. Some participants acknowledged that they had not thought about the ethics or morality of their AI use until interview questions brought them to mind: \textit{``over time, we just do it so much that we don’t really think about if it’s a good idea or not.''} (P6)

Several participants were also unaware of their courses' AI policies. P13 acknowledged they did not know course's policy, admitting: \textit{``To be honest, I didn't think about it.''}.

\subsection{Discussion}
%\jk{1) rationalization critique 2) slippery slope 3) challenges on moral development 4) what can we do about it -- AI policy}

%언니: flow/structure 관련은 rationalization 각각에 대한 혹은 두 개씩 묶어서 critique 하는게 제일 먼저 나오는게 좋을 것 같아요. 하나하나 설명하기보다는 묶어서 설명하는 버전이 더 나아보이고요. 그리고 켄타로에게 쓴 이메일에서 설명한 부분들 좋은 것 같아요. 다만 내용들이 깔끔하게, harm, responsibility, 이런 식으로 의도한대로 떨어지지는 않는 것 같아요. 대신에 그렇게 깔끔하게 나눌 필요 없고, 전체적으로 구분해서 포인트 써도 괜찮아보여요.
%%%Our finding shows college students provided a wide range of  rationalizations for AI use; Some were expressed with little or no explicit discussion of remorse, whereas others were accompanied by significant internal struggle over moral conflicts. This suggests that students' moral reasoning around AI use is not fixed, but remains fluid and unfinished.
Our findings demonstrate 23 distinct rationalizations that university students use to justify their AI use in writing assignments.  When making these claims, some students expressed little or no explicit remorse, while others hinted at internal moral struggles. Students' rationalizations were typically used to explain (unethical) behaviors that had already occurred, rather than being established through moral reasoning or informing their behavior through moral judgment. These post-hoc rationalizations have been reported in the literature on motivated moral reasoning \cite{ditto2009motivated} and social intuitionist model \cite{haidt2001emotional}, implying this type of justification generally lacks normative grounding.

%\jk{Although Draxler et al. \cite{draxler_ai_2024} found that personalization alone did not significantly increase ownership, our findings show that students may still claim ownership when AI-generated text feels aligned with their own style, beliefs, and prior interactions with AI. In particular, some students described “training” ChatGPT to write like them, suggesting that ownership can be shaped by accumulated interaction and perceived voice alignment rather than by one-time personalization alone.}

%Moreover, the way they perceived (ownership of) their writing was contingent. Students tend to consider AI-assisted writing their own when the output sounds like them, or reflects their beliefs or intentions. Investment in shaping the AI output via prompts also made students feel a sense of personal ownership\cite{pierce2001toward}. 

%AI does not merely make academic misconduct easier to perform; it makes questionable AI use easier for students to morally rationalize number of rationaizations simultaneously possible. 
Unlike traditional academic misconduct \citep{maramark1993academic}, AI-assisted writing allows students to reinterpret harm, authorship, responsibility, and contribution in unusually flexible ways. %Because AI is often treated as a tool rather than a human collaborator, students may not experience its use as unauthorized help. 
For example, traditional plagiarism typically involves the victim (original author) being denied \cite{stearns1992copy}. However, with AI, students can claim that no victim exists. At the same time, \textit{Minimal AI Contribution}, \textit{Ex Ante Contribution}, and \textit{Post Hoc Contribution} show how students minimize AI's role while treating their own ideas, prompts, edits, or intentions as decisive evidence of authorship, even though the work delegated to AI may be central to writing-skill development. Together, these dynamics make authorship and responsibility contingent. Students claim ownership when AI output supports credit, but distance themselves when it creates potential risks such as plagiarism or policy violations, as in \textit{Responsibility Denial}, thereby externalizing responsibility in ways that echo moral disengagement \citep{bandura2011moral}. Thus, AI-assisted writing not only challenges conventional understandings of academic misconduct, but also the basic moral principles of authorship, responsibility, and accountability that underlie academic integrity policies. Overall, students' moral reasoning about AI use appears to be fluid, unfinished, and sometimes self-contradictory.

What is also worth noting is that our findings suggest that AI presents a steep, slippery slope\footnote{
Our slippery slope is conceptual, not necessarily temporal, as distinguished by some philosophers \cite{kobis2017road, shalvi2015self}. Slippery slope is often used to describe temporal and/or causal chains of changes in action \cite{volokh2002mechanisms} indeed, some of our participants appeared to have experienced that a minor action with no apparent consequences led to a more serious ethical wrongdoing. However, our emphasis is rather on students' moral status being easily at the very bottom of the slope, relying on the metaphorical concept of it.} of academic integrity for students. %, which we refer to as [ADD our definition/conceptualization]. 
The five sites of AI policy help explain how this slippery slope emerges. Instructor intentions can be translated into formal policies; students then interpret those policies, convert them into personal self-policies, and finally enact them in concrete writing practices. At each site, the boundary of acceptable AI use can be softened or relocated. This slope is interpretive and moral that the meaning of AI support changes as it moves from instructional expectations into students' own reasoning and practices. We emphasize that the slope is slippery: multiple rationalizations exist for every type and degree of AI use -- from fixing minor grammatical errors, all the way to ``vibe-writing'' \cite{jelson2026empirical}, making increasingly extensive AI assistance easier to justify all AI use. And, the slope is very steep: some students reached a moral nadir, submitting writing assignments that contain little of their own writing and claiming them as their own with the rationalizations.  %\jk{This slippery slope is not simply a failure of individual student morality; our finding suggests there is misalignments across faculty intention, formal course policy, student interpretation, student self-policy, and actual practice. These gaps allow students to easily justify departures from one site of policy by appealing to another, making extensive AI delegation may feel morally acceptable.}

Our study findings require attention from the perspectives of faculty and educational institutions as they bring new insights into students' attitudes towards learning (including assignments), writing and ethical practices in educational settings. Students do not appear to view the challenges/limitations to their own learning (caused by using AI) as an ethical issue -- even in a context in which learning is the main goal. Their justifications instead show that goals and processes and tools used for their learning are their own choice. Any formal learning environment arguably involves an implicit social contract in which the instructor helps the student learn, and the student is honest about what they know and do not know \cite{barnhardt2017psychological, gallant2008academic}. However, students' AI-use rationalizations suggest that they are unaware of this latter expectation. %Significantly, few participants mentioned their instructors' difficulties in properly assessing work when the boundary between student and AI work is unclear. 
Significantly, AI use blurs the boundary between student and AI work and it makes it difficult for instructor to assess their work and help them improve their learning. A key intervention for higher education, then, is to help students understand the pedagogical rationale behind AI policies, including why honest representation of one's own knowledge is important for both learning and assessment.

Universities also need to reconsider course assignments -- not just for the sake of assessment, but to help uphold academic integrity standards. %AI can complete many conventional writing assignments in ways that appear acceptable at the level of final products \cite{scarfe2024real, yeadon2023death}. 
Our participants referred to certain assignments as ``Chat-assignments'' suggesting, as reported elsewhere \cite{snipes2023transforming}, that students are well-habituated to thorough AI use. Faculty would do well to design assignments that are ``AI proof,'' but the large-scale loss of the ability to assess homework cannot be borne by faculty alone. Institutional interventions such as, say, ``AI-free study halls'' -- lightly proctored workspaces where students must check in and check out to complete portions of take-home assignments unassisted by digital tools beyond simple word processors -- might be necessary to ensure that at least some take-home assignments are verifiably done by students without AI. 

% \textcolor{red}{Assignments can be redesigned to make students' judgment, or interpretation rather than assigning only a final text.}\jk{i cant think of great recommendation yet...}

%\jk{I want to keep this:} \kt{Edited. Also, I would move this paragraph immediately after the paragraph citing Khalifa. If you move it, replace "Finally," with "Despite the above,".} 
Finally, we do not interpret student rationalizations for AI use as a moral failure on their part, at least not entirely so. Indeed, they express considerable misgivings about the widespread student use of AI. If blame is to be ascribed, it is as at least as much due to the widespread availability of AI tools -- the equivalent of providing primary school students with calculators as they are learning multiplication tables. The problem is arguably as much due to the corporations and universities that provide these powerful tools free to students, with little or no cautions about their hazards, as well as to faculty and departments that have been slow to make provisions for AI-free work. Meanwhile, students are caught in a rapidly changing pedagogical environment, expected to negotiate the boundaries of authorship, responsibility, and acceptable AI use with little else to go on than brief, often out-of-date, instructions in course syllabi. Educational interventions and AI policies should be grounded in cognitively supportive approaches that develop more sustainable forms of AI use (and non-use) in academic contexts, while helping students critically reflect on their behavior \cite{adnin2025examining}. 

One potential application is to use our  rationalizations in Table~\ref{tab:rationalizations-part2} as a basis for student learning and faculty understanding. By making common rationalizations visible, the taxonomy can help students (e.g., reflect their own reasoning) and instructors better understand how students justify AI use, why some justifications may be ethically problematic, and how students might move from unfinished rationalizations toward more deliberate rationales.

\section{Limitations and Future Work}
Our study has several limitations. First, we interviewed 20 undergraduate students who self-reported using AI for writing assignments. Thus, our findings may not represent students who avoid AI, or broader student populations. 
Future work could seek opportunities for studying other populations for whom writing is central, such as researchers, creative writers, journalists, and other writing professionals. Second, we did not interview instructors. As a result, we cannot determine how instructors intended students to interpret or apply their classroom AI policies. Future work could examine instructors' perceptions of AI policy, their pedagogical intentions, and their responses to students' rationalizations. Third, although we aimed to understand students' actual practices by combining interviews with course syllabi, submitted assignments when available, not all participants provided all materials, which may limit our ability to fully capture their behaviors. %Future work could use longitudinal designs to examine how students' rationalizations develop over time and how they change across courses, assignments, and policy contexts.

\section{Conclusion}
%AI seems to blur clear line for today's college students in terms of what is proper and improper, what is moral and not moral as a student.
Through semi-structured interviews with college students in the United States, we present new insights into students' moral sense-making around the use of AI for academic writing. We identify at least five conceptual sites where students negotiate what counts as appropriate AI use, and we show how divergence across these sites contributes to a steep, conceptual slippery slope. Our key findings suggest that there are at least 23 distinct rationalizations that students use to explain why AI use in writing can be ethically acceptable. We also show that some of these rationalizations involve, or are used to justify, violations of course policy, while students also struggle internally with moral conflicts and often emerge post hoc in students' accounts. Overall, our findings suggest that AI policy and educational practice need to do more than establish rules; they should also provide students with cognitive support for navigating ethical ambiguity.

\appendix
%\section{Reference Examples}
\label{sec:reference_examples}

%\section{Acknowledgments}

\bibliography{aaai2026}

@article{draxler2024ai,
  title={The AI ghostwriter effect: When users do not perceive ownership of AI-generated text but self-declare as authors},
  author={Draxler, Fiona and Werner, Anna and Lehmann, Florian and Hoppe, Matthias and Schmidt, Albrecht and Buschek, Daniel and Welsch, Robin},
  journal={ACM Transactions on Computer-Human Interaction},
  volume={31},
  number={2},
  pages={1--40},
  year={2024},
  publisher={ACM New York, NY}
}

@article{clarke2017thematic,
  title={Thematic analysis},
  author={Clarke, Victoria and Braun, Virginia},
  journal={The journal of positive psychology},
  volume={12},
  number={3},
  pages={297--298},
  year={2017},
  publisher={Taylor \& Francis}
}

@article{hawdon2026cheating,
  title={Cheating with ChatGPT and techniques of neutralization},
  author={Hawdon, James and Costello, Matthew and Reichelmann, Ashley V},
  journal={Deviant Behavior},
  volume={47},
  number={4},
  pages={664--685},
  year={2026},
  publisher={Taylor \& Francis}
}

@inproceedings{adnin2025examining,
  title={Examining Student and Teacher Perspectives on Undisclosed Use of Generative AI in Academic Work},
  author={Adnin, Rudaiba and Pandkar, Atharva and Yao, Bingsheng and Wang, Dakuo and Das, Maitraye},
  booktitle={Proceedings of the 2025 CHI Conference on Human Factors in Computing Systems},
  pages={1--17},
  year={2025}
}

@book{eaton2021plagiarism,
  title={Plagiarism in higher education},
  author={Eaton, Sarah Elaine},
  year={2021},
  publisher={Bloomsbury Publishing}
}

@article{wise2024scholarly,
  title={A scholarly dialogue: Writing scholarship, authorship, academic integrity and the challenges of AI},
  author={Wise, Beck and Emerson, Lisa and Van Luyn, Ariella and Dyson, Bronwen and Bjork, Collin and Thomas, Susan E},
  journal={Higher Education Research \& Development},
  volume={43},
  number={3},
  pages={578--590},
  year={2024},
  publisher={Taylor \& Francis}
}

@article{giray2026cheating,
  title={Cheating Writing with Generative AI: Exploring Student Motivations Using the Theory of Planned Behavior},
  author={Giray, Louie and Jacob, Jomarie and Encanto, Valerie and Mansilungan, Crisza Joy},
  journal={Journal of Academic Ethics},
  volume={24},
  number={1},
  pages={1--23},
  year={2026},
  publisher={Springer}
}

@article{fleckenstein2024teachers,
  title={Do teachers spot AI? Evaluating the detectability of AI-generated texts among student essays},
  author={Fleckenstein, Johanna and Meyer, Jennifer and Jansen, Thorben and Keller, Stefan D and K{\"o}ller, Olaf and M{\"o}ller, Jens},
  journal={Computers and Education: Artificial Intelligence},
  volume={6},
  pages={100209},
  year={2024},
  publisher={Elsevier}
}

@article{parker2026longitudinal,
  title={Longitudinal insights into AI in education: Usage, ethics, and policy development in higher education},
  author={Parker, Luke and Loper, A Jane and Carter, Christopher W and Hayes, Josh and Karakas, Alice},
  journal={Computers and Education Open},
  volume={10},
  pages={100329},
  year={2026},
  publisher={Elsevier}
}

@article{emig1977writing,
  title={Writing as a mode of learning},
  author={Emig, Janet},
  journal={College Composition \& Communication},
  volume={28},
  number={2},
  pages={122--128},
  year={1977},
  publisher={NCTE}
}

@article{mccutchen2000knowledge,
  title={Knowledge, processing, and working memory: Implications for a theory of writing},
  author={McCutchen, Deborah},
  journal={Educational psychologist},
  volume={35},
  number={1},
  pages={13--23},
  year={2000},
  publisher={Taylor \& Francis}
}

@article{bandura2011moral,
  title={Moral disengagement},
  author={Bandura, Albert},
  journal={The encyclopedia of peace psychology},
  year={2011},
  publisher={Wiley Online Library}
}

@article{shalvi2015self,
  title={Self-serving justifications: Doing wrong and feeling moral},
  author={Shalvi, Shaul and Gino, Francesca and Barkan, Rachel and Ayal, Shahar},
  journal={Current directions in psychological science},
  volume={24},
  number={2},
  pages={125--130},
  year={2015},
  publisher={Sage Publications Sage CA: Los Angeles, CA}
}

@article{kobis2017road,
  title={The road to bribery and corruption: Slippery slope or steep cliff?},
  author={K{\"o}bis, Nils C and Van Prooijen, Jan-Willem and Righetti, Francesca and Van Lange, Paul AM},
  journal={Psychological science},
  volume={28},
  number={3},
  pages={297--306},
  year={2017},
  publisher={Sage Publications Sage CA: Los Angeles, CA}
}

@article{scarfe2024real,
  title={A real-world test of artificial intelligence infiltration of a university examinations system: A “Turing Test” case study},
  author={Scarfe, Peter and Watcham, Kelly and Clarke, Alasdair and Roesch, Etienne},
  journal={PloS one},
  volume={19},
  number={6},
  pages={e0305354},
  year={2024},
  publisher={Public Library of Science San Francisco, CA USA}
}

@article{barnhardt2017psychological,
  title={Psychological teaching-learning contracts: Academic integrity and moral psychology},
  author={Barnhardt, Bradford and Ginns, Paul},
  journal={Ethics \& Behavior},
  volume={27},
  number={4},
  pages={313--334},
  year={2017},
  publisher={Taylor \& Francis}
}

@article{gallant2008academic,
  title={Academic Integrity in the Twenty-First Century: A Teaching and Learning Imperative. ASHE Higher Education Report, Volume 33, Number 5.},
  author={Gallant, Tricia Bertram},
  journal={ASHE higher education report},
  volume={33},
  number={5},
  pages={1--143},
  year={2008},
  publisher={ERIC}
}

@article{fu2026everyone,
  title={" Everyone's using it, but no one is allowed to talk about it": College Students' Experiences Navigating the Higher Education Environment in a Generative AI World},
  author={Fu, Yue and Lin, Yifan and Wang, Yessica and Tran, Sarah and Hiniker, Alexis},
  journal={arXiv preprint arXiv:2602.17720},
  year={2026}
}

@article{lin2026relying,
  title={Relying on LLMs: Student Practices and Instructor Norms are Changing in Computer Science Education},
  author={Lin, Xinrui and Huang, Heyan and Shi, Shumin and Vines, John},
  journal={arXiv preprint arXiv:2602.05506},
  year={2026}
}

@inproceedings{rismani2026use,
  title={From Use to Oversight: How Mental Models Influence User Behavior and Output in AI Writing Assistants},
  author={Rismani, Shalaleh and Blodgett, Su Lin and Liao, Q Vera and Olteanu, Alexandra and Moon, AJung},
  booktitle={Proceedings of the 2026 CHI Conference on Human Factors in Computing Systems},
  pages={1--23},
  year={2026}
}

@article{ammari2025students,
  title={How students (really) use ChatGPT: Uncovering experiences among undergraduate students},
  author={Ammari, Tawfiq and Chen, Meilun and Zaman, SM and Garimella, Kiran},
  journal={arXiv preprint arXiv:2505.24126},
  year={2025}
}

@book{snipes2023transforming,
  title={Transforming Education With AI: Guide to Understanding and Using ChatGPT in the Classroom},
  author={Snipes, Shane},
  volume={1},
  year={2023},
  publisher={Dr. Shane Snipes, PhD}
}

@article{ditto2009motivated,
  title={Motivated moral reasoning},
  author={Ditto, Peter H and Pizarro, David A and Tannenbaum, David},
  journal={Psychology of learning and motivation},
  volume={50},
  pages={307--338},
  year={2009},
  publisher={Elsevier}
}

@article{haidt2001emotional,
  title={The emotional dog and its rational tail: a social intuitionist approach to moral judgment.},
  author={Haidt, Jonathan},
  journal={Psychological review},
  volume={108},
  number={4},
  pages={814},
  year={2001},
  publisher={American Psychological Association}
}

@article{perkins2023academic,
  title={Academic integrity considerations of AI large language models in the post-pandemic era: ChatGPT and beyond},
  author={Perkins, Mike},
  journal={Journal of University Teaching and Learning Practice},
  volume={20},
  number={2},
  pages={1--24},
  year={2023},
  publisher={Open Access Publishing Association (OAPA) Launceston, Tasmania}
}

@article{cotton2024chatting,
  title={Chatting and cheating: Ensuring academic integrity in the era of ChatGPT},
  author={Cotton, Debby RE and Cotton, Peter A and Shipway, J Reuben},
  journal={Innovations in education and teaching international},
  volume={61},
  number={2},
  pages={228--239},
  year={2024},
  publisher={Taylor \& Francis}
}

@article{eaton2023postplagiarism,
  title={Postplagiarism: Transdisciplinary ethics and integrity in the age of artificial intelligence and neurotechnology},
  author={Eaton, Sarah Elaine},
  journal={International Journal for Educational Integrity},
  volume={19},
  number={1},
  pages={23},
  year={2023},
  publisher={Springer}
}

@article{bittle2025generative,
  title={Generative AI and academic integrity in higher education: A systematic review and research agenda},
  author={Bittle, Kyle and El-Gayar, Omar},
  journal={Information},
  volume={16},
  number={4},
  pages={296},
  year={2025},
  publisher={MDPI}
}

@article{balalle2025reassessing,
  title={Reassessing academic integrity in the age of AI: A systematic literature review on AI and academic integrity},
  author={Balalle, Himendra and Pannilage, Sachini},
  journal={Social Sciences \& Humanities Open},
  volume={11},
  pages={101299},
  year={2025},
  publisher={Elsevier}
}

@article{chan2025students,
  title={Students’ perceptions of ‘AI-giarism’: Investigating changes in understandings of academic misconduct},
  author={Chan, Cecilia Ka Yuk},
  journal={Education and Information Technologies},
  volume={30},
  number={6},
  pages={8087--8108},
  year={2025},
  publisher={Springer}
}

@inproceedings{chakravorti2025social,
  title={Social Scientists on the Role of AI in Research},
  author={Chakravorti, Tatiana and Wang, Xinyu and Venkit, Pranav Narayanan and Koneru, Sai and Munger, Kevin and Rajtmajer, Sarah},
  booktitle={Proceedings of the AAAI/ACM Conference on AI, Ethics, and Society},
  volume={8},
  pages={528--540},
  year={2025}
}

@inproceedings{khosrowi2023diffusing,
  title={Diffusing the creator: Attributing credit for generative AI outputs},
  author={Khosrowi, Donal and Finn, Finola and Clark, Elinor},
  booktitle={Proceedings of the 2023 AAAI/ACM Conference on AI, Ethics, and Society},
  pages={890--900},
  year={2023}
}

@inproceedings{mueller2024llms,
  title={Llms and memorization: On quality and specificity of copyright compliance},
  author={Mueller, Felix B and G{\"o}rge, Rebekka and Bernzen, Anna K and Pirk, Janna C and Poretschkin, Maximilian},
  booktitle={Proceedings of the AAAI/ACM Conference on AI, Ethics, and Society},
  volume={7},
  pages={984--996},
  year={2024}
}

@article{lund2025ai,
  title={AI and academic integrity: Exploring student perceptions and implications for higher education},
  author={Lund, Brady D and Lee, Tae Hee and Mannuru, Nishith Reddy and Arutla, Nikhila},
  journal={Journal of Academic Ethics},
  volume={23},
  number={3},
  pages={1545--1565},
  year={2025},
  publisher={Springer}
}

@article{gruenhagen2024rapid,
  title={The rapid rise of generative AI and its implications for academic integrity: Students’ perceptions and use of chatbots for assistance with assessments},
  author={Gruenhagen, Jan Henrik and Sinclair, Peter M and Carroll, Julie-Anne and Baker, Philip RA and Wilson, Ann and Demant, Daniel},
  journal={Computers and Education: Artificial Intelligence},
  volume={7},
  pages={100273},
  year={2024},
  publisher={Elsevier}
}

@article{waqas2026understanding,
  title={Understanding AIgiarism in higher education: the lens of general AI attitudes and moral disengagement},
  author={Waqas, Muhammad and Hania, Alishba and Chunyan, XU},
  journal={Studies in Higher Education},
  volume={51},
  number={4},
  pages={910--926},
  year={2026},
  publisher={Taylor \& Francis}
}

@article{khlaif2025redesigning,
  title={Redesigning assessments for AI-enhanced learning: A framework for educators in the generative AI era},
  author={Khlaif, Zuheir N and Alkouk, Wejdan Awadallah and Salama, Nisreen and Abu Eideh, Belal},
  journal={Education Sciences},
  volume={15},
  number={2},
  pages={174},
  year={2025},
  publisher={MDPI}
}

@article{ardito2025generative,
  title={Generative AI detection in higher education assessments},
  author={Ardito, Cesare Giulio},
  journal={New Directions for Teaching and Learning},
  volume={2025},
  number={182},
  pages={11--28},
  year={2025},
  publisher={Wiley Online Library}
}

@article{liang2023gpt,
  title={GPT detectors are biased against non-native English writers},
  author={Liang, Weixin and Yuksekgonul, Mert and Mao, Yining and Wu, Eric and Zou, James},
  journal={Patterns},
  volume={4},
  number={7},
  year={2023},
  publisher={Elsevier}
}

@inproceedings{jelson2026empirical,
  title={An empirical study to understand how students use ChatGPT for writing essays},
  author={Jelson, Andrew and Manesh, Daniel and Jang, Alice and Dunlap, Daniel and Kim, Young-Ho and Lee, Sang Won},
  booktitle={Proceedings of the 2026 CHI Conference on Human Factors in Computing Systems},
  pages={1--26},
  year={2026}
}

@article{emsley2023chatgpt,
  title={ChatGPT: these are not hallucinations--they’re fabrications and falsifications},
  author={Emsley, Robin},
  journal={Schizophrenia},
  volume={9},
  number={1},
  pages={52},
  year={2023},
  publisher={Nature Publishing Group UK London}
}

@inproceedings{slimi2026systematic,
  title={A systematic critical review of generative AI's impact on authorship, pedagogy, and integrity (2023--2025)},
  author={Slimi, Zouhaier},
  booktitle={Frontiers in Education},
  volume={11},
  pages={1769680},
  year={2026},
  organization={Frontiers Media SA}
}

@misc{foltynek2023enai,
  title={ENAI Recommendations on the ethical use of Artificial Intelligence in Education},
  author={Foltynek, Tomas and Bjelobaba, Sonja and Glendinning, Irene and Khan, Zeenath Reza and Santos, Rita and Pavletic, Pegi and Kravjar, J{\'u}lius},
  journal={International Journal for Educational Integrity},
  volume={19},
  number={1},
  pages={1--4},
  year={2023},
  publisher={Springer}
}

@article{perkins2024decoding,
  title={Decoding academic integrity policies: A corpus linguistics investigation of AI and other technological threats},
  author={Perkins, Mike and Roe, Jasper},
  journal={Higher Education Policy},
  volume={37},
  number={3},
  pages={633--653},
  year={2024},
  publisher={Springer}
}

@article{perkins2024artificial,
  title={The Artificial Intelligence Assessment Scale (AIAS): A framework for ethical integration of generative AI in educational assessment},
  author={Perkins, Mike and Furze, Leon and Roe, Jasper and MacVaugh, Jason},
  journal={Journal of University Teaching and Learning Practice},
  volume={21},
  number={6},
  pages={49--66},
  year={2024},
  publisher={Open Access Publishing Association (OAPA) Launceston, Tasmania}
}

@article{huang2025academic,
  title={Academic cheating with generative AI: Exploring a moral extension of the theory of planned behavior},
  author={Huang, Dongpeng and Hash, Nicole and Cummings, James J and Prena, Kelsey},
  journal={Computers and Education: Artificial Intelligence},
  volume={8},
  pages={100424},
  year={2025},
  publisher={Elsevier}
}

@article{wu2024ai,
  title={AI governance in higher education: Case studies of guidance at Big Ten universities},
  author={Wu, Chuhao and Zhang, He and Carroll, John M},
  journal={Future Internet},
  volume={16},
  number={10},
  pages={354},
  year={2024},
  publisher={MDPI}
}

@article{theoharakis2025ai,
  title={AI’s learning paradox: how business students’ engagement with AI amplifies moral disengagement-driven misconduct},
  author={Theoharakis, Vasilis and Mylonopoulos, Nikolaos and Papadopoulou, Konstantina},
  journal={Studies in Higher Education},
  pages={1--18},
  year={2025},
  publisher={Taylor \& Francis}
}

@book{maramark1993academic,
  title={Academic dishonesty among college students},
  author={Maramark, Sheilah},
  year={1993},
  publisher={US Department of Education, Office of Educational Research and Improvement~…}
}

@article{stearns1992copy,
  title={Copy wrong: Plagiarism, process, property, and the law},
  author={Stearns, Laurie},
  journal={Calif. L. Rev.},
  volume={80},
  pages={513},
  year={1992},
  publisher={HeinOnline}
}

@article{volokh2002mechanisms,
  title={The mechanisms of the slippery slope},
  author={Volokh, Eugene},
  journal={Harv. L. Rev.},
  volume={116},
  pages={1026},
  year={2002},
  publisher={HeinOnline}
}


\end{document}